\documentclass[twocolumn,prb,superscriptaddress,showpacs]{revtex4-1}
\usepackage{amsmath}
\usepackage{amssymb}
\usepackage{graphicx}
\usepackage{bm}
\usepackage{epstopdf}

\newcommand{\bs}{\mathbf{S}}

\begin{document}

\title{U(1) symmetry of the spin-orbit coupled Hubbard model on the Kagome lattice}
\author{Se Kwon Kim}
\affiliation{Department of Physics and Astronomy, University of California, Los Angeles, California 90095, USA}
\affiliation{Department of Physics and Astronomy, Johns Hopkins University, Baltimore, Maryland 21218, USA}
\author{Jiadong Zang}
\affiliation{Department of Physics and Material Science Program, University of New Hampshire, Durham, New Hampshire 03824, USA}
\affiliation{Department of Physics and Astronomy, Johns Hopkins University, Baltimore, Maryland 21218, USA}
\date{\today}

\begin{abstract}
We theoretically study the symmetry properties of the single-band Hubbard model with general spin-orbit coupling (SOC) on the Kagome lattice. We show that the global U(1) spin-rotational symmetry is present in the Hubbard Hamiltonian owing to the inversion symmetry centered at sites. The corresponding spin Hamiltonian has, therefore, the SO(2) spin-rotational symmetry, which can be captured by including SOC non-perturbatively. The exact classical groundstates, which we obtain for arbitrary SOC, are governed by the SU(2) fluxes associated with SOC threading the constituent triangles. The groundstates break the SO(2) symmetry, and the associated Berezinsky-Kosterlitz-Thouless transition temperature is determined by the SU(2) fluxes through the triangles, which we confirm by finite temperature classical Monte Carlo simulation.
\end{abstract}


\maketitle

\section{Introduction}

The Hubbard model \cite{HubbardPRSLA1963} has been acknowledged as a paradigm of strongly correlated electron systems. Despite its simplicity, the Hubbard model can exhibit various phenomena from antiferromagnetism \cite{AndersonSSP1963} to metal-insulator transition \cite{MottRMP1968} and high-temperature superconductivity \cite{ScalapinoJSNM2006}. The Hubbard model without spin-orbit coupling (SOC) has nontrivial symmetry properties besides the apparent global SU(2) spin-rotational symmetry, which has guided us to the exploration of rich phases and excitations \cite{AffleckPRB1988, YangPRL1989, ZhangPRL1990, HermelePRB2007}. \textcite{YangMPLB1990}, for example, revealed the SU(2) pseudo spin symmetry, which includes the U(1) phase symmetry as a subgroup, and predicted the massive collective modes in any phase-symmetry-breaking superconductivity.

In spin systems, SOC gives rise to frustration on spin interactions and reduces the symmetry in general. \cite{MoriyaPR1960} The spin-dependent hopping, the manifestation of SOC in the kinetic terms of the Hubbard model, can be described by an SU(2) gauge field. \cite{KaplanZPB1983, ShekhtmanPRL1992, GuarnacciaPRB2012, ZhuPRB2014} In open-ended one-dimensional chains, the SU(2) field can be gauged away by a string of gauge transformations \cite{KaplanZPB1983}, wherein the global SU(2) spin-rotational symmetry is intact. In rings, however, the SU(2) field creates a nonvanishing flux in general, which makes the system frustrated and reduces the symmetry down to U(1). \cite{ShekhtmanPRL1992} SU(2) symmetry is recovered only when the enclosed SU(2) flux vanishes. \cite{MeirPRL1989, TserkovnyakPRB2007} Two-dimensional lattices are composed of interconnected loops, each of which embraces the flux. General SOC breaks the continuous symmetry, \footnote{A few exceptions are known, e.g., the continuous symmetry persists when the magnitudes of the Rashba and Dresselhaus coupling constants are equal \protect \cite{GuarnacciaPRB2012}.} and engenders a long-ranged magnetic order escaping the Mermin-Wagner theorem \cite{MerminPRL1966}.

The geometry of the lattice is another source of frustration. The Kagome lattice, a two-dimensional lattice of corner-sharing triangles, is a prototypical example that brings geometric frustration to antiferromagnetic materials exemplified by herbertsmithite ZnCu$_3$(OH)$_6$Cl$_2$. The Kagome lattice Hubbard model without SOC has been extensively studied in metal-insulator transitions \cite{OhashiPRL2006, KurataniJPCM2007, BernhardJPCM2007} and the van Hove filling \cite{YuPRB2012, KieselPRL2013, WangPRB2013}. The corresponding spin Hamiltonian of the Hubbard model in the large-$U$ limit at half filling has been studied in search of exotic phases on the Kagome lattice, such as spin liquids \cite{Lacroix2011}. The physical effects of SOC in the spin Hamiltonian have been studied by including its leading order contribution to the Hamiltonian known as the Dzyaloshinskii-Moriya (DM) interaction \cite{DzyaloshinskiiJPCS1958, MoriyaPR1960}, which has been known to induce a long-ranged magnetic order. \cite{ElhajalPRB2002, RigolPRL2007, HermelePRB2007, CepasPRB2008, HuhPRB2010, TovarPRB2009, ZorkoPRL2011, HwangPRB2012}

In this paper, we show that the global U(1) spin-rotational symmetry is present in the single-band SOC Hubbard Hamiltonian on the Kagome lattice owing to the inversion symmetry centered at sites \cite{RigolPRL2007}. The corresponding spin Hamiltonian has, therefore, the SO(2) spin-rotational symmetry, which can be captured by including SOC non-perturbatively. The exact classical groundstates, which we obtain for arbitrary SOC, are governed by the SU(2) fluxes associated with SOC threading the constituent triangles. The groundstates break the continuous symmetry, and the associated Berezinsky-Kosterlitz-Thouless (BKT) transition temperature is determined by the SU(2) fluxes through the triangles, which we confirm by finite temperature classical Monte Carlo simulation. \footnote{On the Kagome lattice, it has been known that DM interaction (a leading order contribution from SOC to the spin Hamiltonian) can destroy the continuous spin symmetry and induce a long-ranged magnetic order, escaping the Mermin-Wagner theorem \cite{RigolPRL2007, ZorkoPRL2011}. The conclusion changes if we consider the full SOC as done in this paper. The U(1) continuous symmetry exists even in the presence of the SOC on the Kagome lattice and the classical groundstates break that symmetry, meaning that SOC cannot cause a long-ranged magnetic order.}

\begin{figure}
\includegraphics[width=0.95\columnwidth]{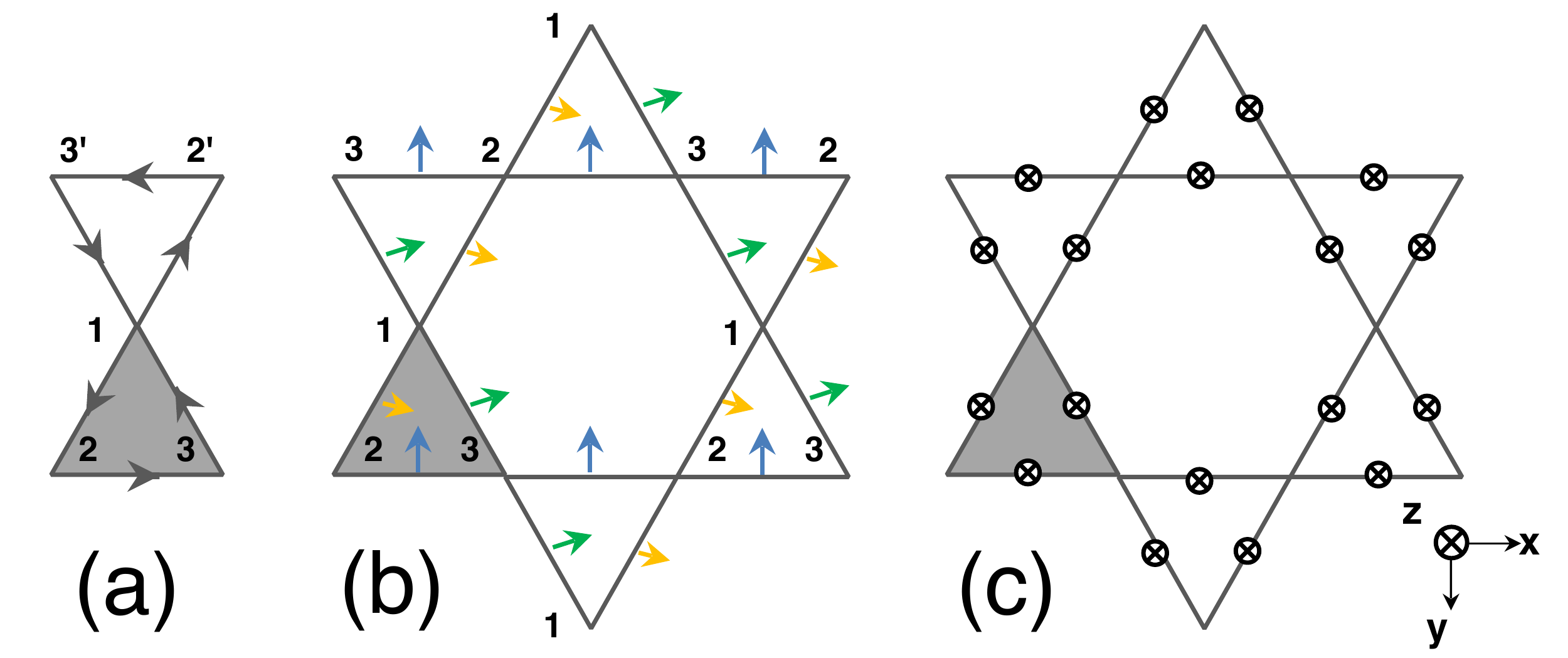}
\caption{(Color online) (a) The site labeling scheme and the direction of the links, $j \rightarrow k$, of the DM vectors $\mathbf{d}_{jk}$. (b), (c) The DM vector $\mathbf{d}_{jk}$ of the gauge fields $U_{jk} \equiv \exp(- i \mathbf{d}_{jk} \cdot \boldsymbol{\sigma} / 2)$ on the link $j \rightarrow k$, directed counterclockwise around triangles (b) in the original Hubbard Hamiltonian $\hat{H}_1$ and (c) in the gauge-transformed Hamiltonian $\hat{H}_2$ (for $\hat{\mathbf{n}} = \hat{\mathbf{z}}$), respectively. }
\label{fig:fig1}
\end{figure}

\section{Summary of main results}

We study the the single-band Hubbard model to describe SOC electron systems on the lattice \footnote{We chose to study the Hubbard model because it is one of the simplest models for description of electronic systems that allow us to incorporate the effects of SOC and lattice structure effectively. Physical properties of herbertsmithite are known to be largely governed by these two effects \cite{RigolPRL2007, CepasPRB2008, GuertlerPRB2014}.}:
\begin{equation}
\hat{H}_1 \equiv - t \sum_{\langle j, k \rangle} \hat{c}_j^\dagger U_{jk}  \hat{c}_k + U \sum_j \hat{n}_{j \uparrow} \hat{n}_{j \downarrow} \, ,
\label{eq:H1}
\end{equation}
where $\langle j, k \rangle$ represents the nearest neighbors $j$ and $k$, $\hat{c}^\dagger_j \equiv (\hat{c}^\dagger_{j \uparrow}, \hat{c}^\dagger_{j \downarrow})$ and $\hat{c}_j \equiv (\hat{c}_{j \uparrow}, \hat{c}_{j \downarrow})^\mathrm{T}$ are the electron creation and annihilation operators, and $\hat{n}_{j \alpha} \equiv \hat{c}^\dagger_{j \alpha} \hat{c}_{j \alpha}$ is the electron number operator with the spin $\alpha$. Here, $t$ is a real hopping magnitude \footnote{The hopping magnitude $t$ may depend on the link. It, however, does not affect our main result, the U(1) symmetry of the Hubbard Hamiltonian.};
\begin{equation}
U_{jk} \equiv \exp(- i \mathbf{d}_{jk} \cdot \boldsymbol{\sigma} / 2)
\end{equation}
describes the effect of SOC, which rotates spin of an electron while hopping \footnote{The form of the hopping matrix, $- t U_{jk}$, is dictated by the invariance of the Hamiltonian $\hat{H}_1$ under the time reversal, $\hat{T} \equiv \hat{K} \prod_j e^{-i \pi \hat{S}_j^y}$, where $\hat{K}$ is the complex conjugate operator and $\hat{\mathbf{S}}_j \equiv \hat{c}_j^\dagger (\boldsymbol{\sigma} / 2) \hat{c}_j$ is the spin operator at the site $i$ \cite{ZanonJPF1988}.}; $\hat{\mathbf{d}}_{jk}$ is the direction of the DM vector; $\boldsymbol{\sigma}$ is the vector of Pauli matrices; $U$ is the magnitude of the on-site Coulomb repulsion. Hermicity of the Hamiltonian requires $U_{jk} = U_{kj}^\dagger$, and thus $\mathbf{d}_{jk} = - \mathbf{d}_{kj}$.

The DM vectors $\mathbf{d}_{jk}$ are physical, but can be considered as a particular realization of the SU(2) gauge field in the lattice gauge theory \cite{GuarnacciaPRB2012}, which provides a suitable language to study the symmetry of the Hubbard Hamiltonian. The local SU(2) gauge transformation, $\hat{c}_j \mapsto V_j \hat{c}_j$ and $U_{jk} \mapsto V_j U_{jk}  V_k^\dagger$, corresponds to the rotation of local spin axes. 

The symmetry of the Hubbard Hamiltonian with SOC is closely related to the SU(2) flux vector $\boldsymbol{\Phi}$ enclosed by loops on the lattice. \cite{ShekhtmanPRL1992, LiNJP2015} It is defined by
\begin{equation}
\exp(-i \boldsymbol{\Phi} \cdot \boldsymbol{\sigma} / 2) \equiv \prod_{j \rightarrow k} U_{jk}
\end{equation}
for each loop, where $j \rightarrow k$ means that sites are traversed counterclockwise around the loop as shown in Fig.~\ref{fig:fig1}(a).  In the absence of SOC,  the fluxes vanish, which results in the invariance of the Hamiltonian under the global SU(2) spin rotation, $\hat{c}_j \mapsto V \hat{c}_j$. A finite SOC causes nontrivial fluxes through loops, which would reduce the symmetry from the continuous SU(2) to the discrete Z$_2$, $\hat{c}_j \mapsto - \hat{c}_j$, generally.

The continuous symmetry, however, can persist even when SOC is present, if the SU(2) fluxes through the loops meet a certain condition. Our main discovery is the U(1) spin-rotational symmetry of the Kagome lattice Hubbard model with SOC. In the Kagome lattice, the inversion symmetry centered at sites \cite{RigolPRL2007} demands matching of the DM vectors of adjacent triangles. As a result all the triangles share the same SU(2) flux structure, $\exp(- i \boldsymbol{\Phi} \cdot \boldsymbol{\sigma} / 2) \equiv U_{12} U_{23} U_{31}$, with the site labeling as shown in Fig.~\ref{fig:fig1}(a) and \ref{fig:fig1}(b), where $\boldsymbol{\Phi} \equiv \Phi \hat{\mathbf{n}}$ is the SU(2) flux vector threading the triangles. For an isolated triangle, the Hubbard Hamiltonian with arbitrary SOC possesses the global U(1) spin-rotational symmetry, and it can be revealed by the local SU(2) gauge transformations that are determined by the SU(2) flux through the triangle. \cite{KaplanZPB1983} Sharing the same SU(2) flux between the adjacent triangles in the Kagome lattice extends the U(1) symmetry of an isolated triangle to the entire lattice, which becomes visible in the gauge-transformed Hamiltonian,
\begin{equation}
\hat{H}_2 \equiv -t \sum_{j \rightarrow k} [\hat{c}'^\dagger_j e^{- i \phi \hat{\mathbf{n}} \cdot \boldsymbol{\sigma} / 2} \hat{c}'_k + \text{H.c.}] + U \sum_j \hat{n}'_{j \uparrow} \hat{n}'_{j \downarrow} \, ,
\label{eq:H2}
\end{equation}
characterized by the single DM vector $\phi \hat{\mathbf{n}}$ [Fig.~\ref{fig:fig1}(c)], where $\hat{c}'_j = U_j \hat{c}_j$ is the new electron operator and $U_j$ describes the local gauge transformation that is governed by the SU(2) flux. The angle $\phi$ is uniquely determined up to $2 \pi / 3$ by the SU(2) flux through the triangle, $3 \phi = \Phi$ mod $2 \pi$.

\begin{figure}
\includegraphics[width=0.95\columnwidth]{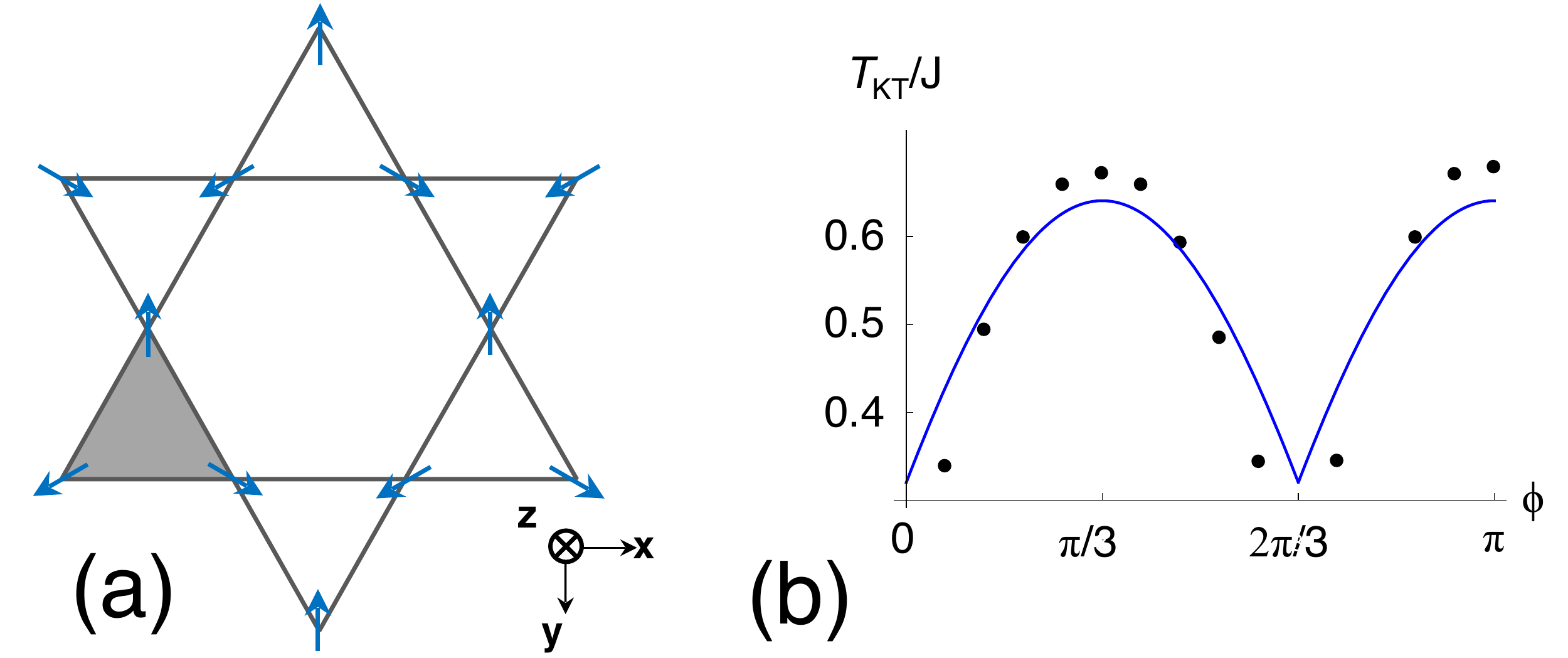}
\caption{(Color online) (a) One of the exact groundstates of the spin Hamiltonian $H_2^s$~(\ref{eq:Hs2}) for the angle $0 < \phi < 2 \pi / 3$ and $\hat{\mathbf{n}} = \hat{\mathbf{z}}$. Any global spin rotation of the state about the axis $\hat{\mathbf{n}}$ yields also a groundstate. (b) BKT transition temperature of $H^s_2$ as a function of the angle $\phi$. The transition temperature $T_\text{KT}$ is modulated by the flux through triangles, $\Phi = 3 \phi$. Dots: finite temperature Monte Carlo simulation results. Line: theoretical result $T_\text{KT}$~(\ref{eq:T_KT}) with the renormalized spin length $S = 0.69$.}
\label{fig:fig2}
\end{figure}

The continuous symmetry also manifests itself in the corresponding spin model,
\begin{equation}
H^s_1 \equiv J \sum_{j \rightarrow k} \bs_j \cdot R (\mathbf{d}_{jk}) \cdot \bs_k \, ,
\label{eq:Hs1}
\end{equation}
that is derived from the large-$U$ limit of the Hubbard Hamiltonian $\hat{H}_1$~(\ref{eq:H1}) at half filling. Here $J \equiv 8 t^2 / U$ sets the energy scale of spin interactions; $R (\mathbf{d}_{jk})$ is the SO(3) matrix of a rotation about the axis $\hat{\mathbf{d}}_{jk}$ with the angle $\phi_{jk} \equiv |\mathbf{d}_{jk}|$. The summand can be split into three terms with the aid of Rodrigues' rotation formula:
\begin{equation}
\begin{split}
 J \bs_j \cdot R (\mathbf{d}_{jk}) \cdot \bs_k =
	& J \cos \phi_{jk} \, \bs_j \cdot \bs_k \\
	& - J \sin \phi_{jk} \, \hat{\mathbf{d}}_{jk} \cdot \bs_j \times \bs_k \\
	& + J(1 - \cos \phi_{jk}) \, (\bs_j \cdot \hat{\mathbf{d}}_{jk}) (\bs_k \cdot \hat{\mathbf{d}}_{jk}) \, . \nonumber
\end{split}
\end{equation}
The first term is the antiferromagnetic Heisenberg interaction; the second term is the DM interaction, which is antisymmetric in exchanging two spins; the third term is the symmetric anisotropic interaction which always accompanies the DM interaction in insulators. \cite{KaplanZPB1983, ShekhtmanPRL1992, ZheludevPRL1998} Application to the spin Hamiltonian $H_1^s$ of the SO(3) equivalent of the SU(2) gauge transformation connecting $\hat{H}_1$ and $\hat{H}_2$ yields the new spin Hamiltonian,
\begin{equation}
H^s_2 \equiv J \sum_{j \rightarrow k} \bs'_j \cdot R (\phi \hat{\mathbf{n}}) \cdot \bs'_k \, ,
\label{eq:Hs2}
\end{equation}
which has the transparent global SO(2) spin-rotational symmetry that is possible to obtain only if we treat the SOC non-perturbatively.

We shall provide the exact classical groundstates of the spin Hamiltonians $H^s_2$~(\ref{eq:Hs2}) later, from which the groundstates of $H^s_1$~(\ref{eq:Hs1}) can be obtained by reversing the gauge transformation. The groundstates of an isolated triangle is obtained by the exact diagonalization of the Hamiltonian. Inversion symmetry centered at sites makes the groundstates of adjacent triangles compatible, which allows us to minimize the spin energy on the entire Kagome lattice. These groundstates break the SO(2) symmetry of the spin Hamiltonian. We confirm that the associated BKT transition occurs at a finite temperature $T_\mathrm{KT}$ modulated by the flux $3 \phi$ through the triangle with the aid of finite temperature Monte Carlo simulation as shown in Fig.~\ref{fig:fig2}(b).

\section{Symmetry of the SOC Hubbard model}

\subsection{Symmetry on the 1D lattice}

For an open-ended one-dimensional chain $\{ \hat{c}_1, \hat{c}_2, \cdots, \hat{c}_N \}$, one can keep the first electron $\hat{c}_1$ unchanged, and perform a string of successive SU(2) gauge transformations on the rest electrons by
\begin{equation}
\hat{c}'_j \equiv \left( \prod_{1 \le k < j} U_{k,k+1} \right) \hat{c}_j \, ,
\end{equation}
which transforms the original Hubbard Hamiltonian $\hat{H}_1$~(\ref{eq:H1}) to
\begin{equation}
\hat{H} = \sum_j [ -t \hat{c}'^\dagger_j \hat{c}'_{j+1} + \text{H.c.}] +  U \sum_j \hat{n}'_{j \uparrow} \hat{n}'_{j \downarrow} \, ,
\label{eq:H3}
\end{equation}
where the global SU(2) spin-rotational symmetry is evident. \cite{KaplanZPB1983, ShekhtmanPRL1992}

For a one-dimensional ring, the SU(2) symmetry is not present generally because of nontrivial SU(2) flux penetrating the ring. The Wilson line for the ring is the product of the link gauge fields, which is given by
\begin{equation}
e^{- i \mathbf{\Phi} \cdot \boldsymbol{\sigma} / 2} \equiv \prod_{1 \le j \le N+1} U_{j, j+1}
\end{equation}
with the periodic boundary condition $c_{N+1} = c_1$ assumed, where $\mathbf{\Phi} \equiv \Phi \hat{\mathbf{n}}$ is the SU(2) flux vector. The flux magnitude $\Phi$ is uniquely defined up to $2 \pi$ \footnote{$\Phi \mapsto 4 \pi - \Phi$ is equivalent to inversion of the direction $\hat{\mathbf{n}}$. The direction $\hat{\mathbf{n}}$ of the SU(2) flux vector is gauge-dependent, whereas the flux magnitude $\Phi$ is not. Energetic quantities thus depend only on $\Phi$, but not $\hat{\mathbf{n}}$. Spin-related quantities, however, may depend on $\hat{\mathbf{n}}$ as well as $\Phi$.}, and the global SU(2) symmetry is restored only when the flux vanishes, $\Phi = 0$ (mod $2 \pi$). \cite{ShekhtmanPRL1992} The flux vector $\mathbf{\Phi}$ can be evenly distributed to all links by the local SU(2) gauge transformation, which is given by
\begin{equation}
\hat{c}'_j \equiv e^{i (j - 1) \mathbf{\Phi} \cdot \boldsymbol{\sigma} / 2 N} \left( \prod_{1 \le k < j} U_{k, k+1} \right) \hat{c}_j \, .
\label{eq:gt-ring}
\end{equation}
The resultant Hamiltonian is $\hat{H}_2$~(\ref{eq:H2}) with $\phi \equiv \Phi / N$, which is invariant under global U(1) spin rotation about the axis $\hat{\mathbf{n}}$, $\hat{c}'_j \mapsto e^{- i \theta \hat{\mathbf{n}} \cdot \boldsymbol{\sigma} / 2} \hat{c}'_j$ for any angle $\theta$. \footnote{The exact quantum groundstate of the Hamiltonian $\hat{H}_2$~(\ref{eq:H2}) can be obtained by the Bethe ansatz with a twisted boundary condition. \cite{ShastryPRL1990, ZvyaginPRB2013}}

\subsection{Symmetry on the Kagome lattice}

We show that the Hubbard Hamiltonian on the Kagome lattice is invariant under the global U(1) spin rotation, which is protected by the inversion symmetry centered at sites that are respected in many materials such as herbertsmithite ZnCu$_3$(OH)$_6$Cl$_2$ and Fe jarosite compound KFe$_3$(SO$_4$)$_2$(OH)$_6$. \cite{YildirimPRB2006, RigolPRL2007, HermelePRB2007} The SU(2) link gauge field $U_{jk} \equiv \exp(-i \mathbf{d}_{jk} \cdot \boldsymbol{\sigma} / 2)$ can be attributed to the electrostatic potential $V(\mathbf{r})$ induced by surrounding molecules. \cite{AharonovPRL1984, GoldhaberPRL1989} The DM vector $\mathbf{d}_{jk} \propto \boldsymbol{\nabla} V \times (\mathbf{r}_j - \mathbf{r}_k)$ is invariant under the inversion $\mathbf{r} \mapsto - \mathbf{r}$ centered at sites provided that $V(\mathbf{r})$ is even under the inversion. For example, in Fig.~\ref{fig:fig1}(a) and (b), under the inversion centered at the site $1$, the site $2$ (operator $\hat{c}_2$) maps to the site $2'$ (operator $\hat{c}_{2'}$), which transforms the associated kinetic term: 
\begin{equation}
-t \hat{c}_1 \exp(-i \mathbf{d}_{12} \cdot \boldsymbol{\sigma}) \hat{c}_2 \rightarrow -t \hat{c}_1 \exp(-i \mathbf{d}_{12} \cdot \boldsymbol{\sigma}) \hat{c}_{2'} \, .
\end{equation}
The invariance of the Hamiltonian under the transformation requires that the right side of the equation is equivalent to $-t \hat{c}_1 \exp(-i \mathbf{d}_{12'} \cdot \boldsymbol{\sigma}) \hat{c}_{2'}$, and, thus, dictates $\mathbf{d}_{12} = \mathbf{d}_{12'}$.

Each site of the Kagome lattice can be labeled by three numbers $1, 2,$ or $3$, as illustrated in Fig.~\ref{fig:fig1}(a). Once the link gauge fields $U_{12}, U_{23}$, and $U_{31}$ of an arbitrarily-chosen triangle (e.g., a shaded one in Fig.~\ref{fig:fig1}) are fixed, all the other link gauge fields on the Kagome lattice are determined by the inversion symmetry. Since all the triangles have the same SU(2) link gauge fields, the gauge transformations (\ref{eq:gt-ring}) for neighboring triangles are compatible. Specifically, the gauge transformation
\begin{equation}
\hat{c}'_j \equiv
\begin{cases}
	\hat{c}_j \, , 	&\text{if $j$ is labelled by $1$} \\
	e^{i \mathbf{\Phi} \cdot \boldsymbol{\sigma} / 6} U_{12} \hat{c}_j \, , &\text{if $j$ is labelled by $2$} \\
	e^{i \mathbf{\Phi} \cdot \boldsymbol{\sigma} / 3} U_{12} U_{23} \hat{c}_j \, , &\text{if $j$ is labelled by $3$} \\
\end{cases}
\label{eq:gt}
\end{equation}
on the original Hubbard Hamiltonian $\hat{H}_1$~(\ref{eq:H1}) results in the new Hubbard Hamiltonian $\hat{H}_2$~(\ref{eq:H2}) that shows the global U(1) spin-rotational symmetry clearly. \footnote{The conserved quantity associated with the U(1) symmetry of the original Hubbard Hamiltonian $\hat{H}_1$~(\ref{eq:H1}) is $\sum_{j \text{ labelled by 1}} \hat{\mathbf{n}} \cdot \bs_j + \sum_{j \text{ labelled by 2}} \hat{\mathbf{n}} \cdot R(\mathbf{d}_{12}) \bs_j + \sum_{j \text{ labelled by 3}} \hat{\mathbf{n}} \cdot R(\mathbf{d}_{12}) R(\mathbf{d}_{23}) \bs_j)$.} Fig.~\ref{fig:fig1}(b) and (c) show the DM vectors in the original Hamlitonian $\hat{H}_1$~(\ref{eq:H1}) and the gauge-transformed Hamiltonian $\hat{H}_2$~(\ref{eq:H2}) (with $\hat{\mathbf{n}} = \hat{\mathbf{z}}$), respectively.

\section{Groundstates of the SOC spin Hamiltonian}

Starting from the Hubbard Hamiltonian $\hat{H}_2$~(\ref{eq:H1}), the large $U$ limit at half filling freezes the charge fluctuation, and eventually ends up with the spin Hamiltonian $H_1^s$~(\ref{eq:Hs1}) on the second order perturbation in $t/U$. \cite{MoriyaPR1960, ShekhtmanPRL1992} The SO(3) counterpart of the SU(2) gauge transformation in Eq.~(\ref{eq:gt}), given by
\begin{equation}
\bs'_j \equiv
\begin{cases}
	\bs_j, 	&\text{if $j$ is labelled by $1$} \\
	R (- \phi \hat{\mathbf{n}}) R(\mathbf{d}_{12}) \bs_j, &\text{if $j$ is labelled by $2$} \\
	R (- 2 \phi \hat{\mathbf{n}}) R(\mathbf{d}_{12}) R(\mathbf{d}_{23}) \bs_j, &\text{if $j$ is labelled by $3$} \\
\end{cases},
\label{eq:so3-gt}
\end{equation}
yields the new spin Hamiltonian $H_2^s$~(\ref{eq:Hs2}). The spin Hamiltonian $H_2^s$ is invariant under global SO(2) spin rotation about the axis $\hat{\mathbf{n}}$, $\bs'_j \mapsto \mathbf{R} (\theta \hat{\mathbf{n}}) \cdot \bs'_j$, which is the consequence of the U(1) spin-rotational symmetry of the parent Hubbard Hamiltonian. The full SO(3) spin-rotational symmetry is respected once the the flux vanishes $\Phi=0$ (mod $2\pi$), or equivalently, $\phi=0, 2\pi/3$, or $4\pi/3$. \footnote{The angles $\phi$ and $\phi + 2 \pi/3$ are connected by the gauge transformation. Specifically, the gauge transformation, [$\bs'_j \mapsto \bs'_j$ if $j$ is labeled by $1$, $\bs'_j \mapsto R(2 \pi /3 \hat{\mathbf{n}}) \bs'_j$ if $j$ is labeled by $2$, $\bs'_j \mapsto R(4 \pi /3 \hat{\mathbf{n}}) \bs'_j$ if $j$ is labeled by $3$], changes $\phi$ to $\phi + 2 \pi / 3$ in the spin Hamiltonian $H_2^s$~(\ref{eq:Hs2}).}

The dependence of the symmetry of the Hamiltonian on its parameter, $\phi$ in our case, indicates the possible dramatic change in the physical properties when $\phi$ crossing the high symmetry points $\phi=0, 2\pi/3$, and $4\pi/3$. To see that, we treat spins classically in the spin Hamiltonian $H_2^s$~(\ref{eq:Hs2}), \cite{HusePRB1992, ElhajalPRB2002, HenleyPRB2009} which allows us to obtain the exact groundstates. The groundstates at the high symmetry points, where the fluxes are zero, are already known. \cite{ChalkerPRL1992, YildirimPRB2006} We thus focus on nonvanishing fluxes. 

We start for an isolated triangle. Observing the Hamiltonian is quadratic in spin, a straightforward way is to list all three spins into a large column spin $\mathcal{S} \equiv (\bs_1, \bs_2, \bs_3)$, and diagonalize a $9 \times 9$ matrix $\mathcal{H}$ representing the spin Hamiltonian $H_2^s = \mathcal{S} \cdot \mathcal{H} \cdot \mathcal{S} / 2$. Specifically the matrix $\mathcal{H}$ is given by
\begin{equation}
\mathcal{H} \equiv 
\begin{pmatrix}
	0 & R & R^\mathrm{T} \\
	R^\mathrm{T} & 0 & R \\
	R & R^\mathrm{T} & 0
\end{pmatrix} \, ,
\end{equation}
where $R \equiv R(\phi \hat{\mathbf{n}})$. This scheme is generally invalid as the eigenstates may not be physical due to the different lengths of spins, e.g., $|\bs_1| \ne |\bs_2|$. For the current problem, however, the eigenvectors with the minimum energy turn out to satisfy $|\bs_1| = |\bs_2| = |\bs_3|$ always, which makes them physical. Spins are perpendicular to the SU(2) flux vector $\Phi \hat{\mathbf{n}}$ in the continuously degenerate groundstates, which are given by
\begin{equation}
\label{eq:gs}
\begin{cases}
\bs_1 = R \left( \frac{4 \pi \hat{\mathbf{n}}}{3} \right) \bs_2 = R \left( \frac{2 \pi \hat{\mathbf{n}}}{3} \right) \bs_3,	& 0 < \phi < \frac{2 \pi}{3} \\
\mathbf{S}_{1}=\mathbf{S}_{2}=\mathbf{S}_{3},	& \frac{2 \pi}{3} < \phi < \frac{4 \pi}{3} \\
\bs_1 = R \left( \frac{2 \pi \hat{\mathbf{n}}}{3} \right)\bs_2 = R \left( \frac{4 \pi \hat{\mathbf{n}}}{3} \right) \bs_3, 	& \frac{2 \pi}{3} < \phi < 2 \pi
\end{cases}.
\end{equation}

For two neighboring triangles, their exact groundstates of the spin Hamiltonian can be patched by matching the spin of the shared site. This procedure can be extended to the entire Kagome lattice because all the triangles have the same DM vectors. Specifically, spins labeled by the same number [see Fig.~\ref{fig:fig1}(a)] point in the same direction in the groundstates. For example, the groundstates of the spin Hamiltonian $H_2^s$ with $2 \pi / 3 < \phi < 4 \pi /3$ have all the spins pointing in the same direction in the plane perpendicular to $\hat{\mathbf{n}}$. Fig.~\ref{fig:fig2}(a) shows a groundstate of $H_2^s$ for the angle $0 < \phi < 2 \pi / 3$ and $\hat{\mathbf{n}} = \hat{\mathbf{z}}$. The exact groundstates of the spin Hamiltonian $H^s_1$~(\ref{eq:Hs1}) can be obtained from those of $H_2^s$~(\ref{eq:gs}) by reversing the gauge transformation in Eq.~(\ref{eq:so3-gt}).

The groundstates of the spin Hamiltonian break the SO(2) spin-rotational symmetry, which signals the existence of the BKT transition at finite temperature. In the continuum approximation, the transition temperature is given by
\begin{equation}
T_\mathrm{KT} = \left( \pi \sqrt{3} / 4 \right) S^2 \cos(\tilde{\Phi} / 3) \, ,
\label{eq:T_KT}
\end{equation}
where $\tilde{\Phi} = \Phi \text{ mod } 2 \pi$, $2 \pi < \tilde{\Phi}  < 4 \pi$, when neglecting spin waves. Fig.~\ref{fig:fig2}(b) shows the results for the transition temperatures from finite temperature classical Monte Carlo simulation of the spin Hamiltonian $H_2^s$ and various $\phi$ with the spin length of unity, which agrees well with the theoretical prediction (\ref{eq:T_KT}) with a renormalized spin length $S = 0.69$. The renormalization of the spin length can be attributed to thermal spin-wave fluctuations.

\section{Discussion}

We have showed that the single-band Hubbard Hamiltonian on the Kagome has the global U(1) spin-rotational symmetry even in the presence of SOC. The U(1) symmetry does not demand a specific shape of constituent triangles, but only requires the inversion symmetry between neighboring triangles centered at the shared site. Linear deformation of the Kagome lattice by strain, for example, would break the three-fold rotational symmetry about the centers of the triangles, but preserves the inversion symmetry centered at sites and thus maintains the associated U(1) symmetry as well \footnote{The breaking of the three-fold rotational symmetry, however, would give rise to the link-dependent hopping magnitude $t \rightarrow t_{jk}$ in the Hubbard Hamiltonian $\hat{H}_1$. This in turn would make the exchange constant in the spin Hamiltonian $H_1^s$ depend on the link $J \rightarrow J_{jk}$, for which the states given in Eq.~(\ref{eq:gs}) may not be the groundstates.}.

We have provided the exact classical groundstates of the spin Hamiltonian, which spontaneously break the continuous symmetry. The BKT transition occurs at finite temperature, which is governed by the SU(2) flux threading the triangles. This is an example showing the physical effects of the SU(2) flux associated with SOC. Its effect on the quantum Hamiltonian would deserve to be investigated. We would like to mention that the controllable SU(2) gauge field has been created in optical lattices, \cite{OsterlohPRL2005, DalibardRMP2011} which may afford the platform to observe the effects of varying flux on the physical properties of the quantum Hamiltonian.

Preservation of U(1) spin rotational symmetry depends on lattice structure. The Kagome lattice (with inversion symmetry centered at sites) is one of the lattices whose structure support U(1) spin rotational symmetry even in the presence of arbitrary SOC. The approach taken in this paper, however, can be applied to other lattice systems. For example, on the square lattice, the Hubbard Hamiltonian with arbitrary SOC  also possesses the global U(1) spin-rotational symmetry provided the inversion symmetry centered at sites is respected. The classical groundstates of the corresponding spin Hamiltonian are the N\'eel states polarized along the SU(2) flux vector, for the square lattice is bipartite. The groundstates do not break the continuous symmetry, and thus the BKT transition does not occur. It would be worth pursuing to study other two- and three-dimensional lattices in the similar approach.

\begin{acknowledgments}
We are grateful to Scott Bender, So Takei, and Oleg Tchernyshyov for useful comments on the manuscript, and appreciate the insightful discussions with Naoto Nagaosa, Masaki Oshikawa, Ian Spielman, and Yuan Wan. This work was supported in part by U.S. Department of Energy, Office of Basic Energy Sciences under Award No. DE-FG02-08ER46544 (S.K.K. and J.Z.) and $\text{DE-SC0012190}$ (S.K.K.).
\end{acknowledgments}

\bibliographystyle{/Users/evol/Dropbox/School/Research/apsrev4-1-nourl}
\bibliography{/Users/evol/Dropbox/School/Research/master}

\end{document}